

\def\boxx{\bar \sqcup}
\headline={\ifnum\pageno=1\firstheadline\else
\ifodd\pageno\rightheadline \else\leftheadline\fi\fi}
\def\firstheadline{\hfil}
\def\rightheadline{\hfil}
\def\leftheadline{\hfil}
	\footline={\ifnum\pageno=1\firstfootline\else\otherfootline\fi}
\def\firstfootline{\rm\hss\folio\hss}
\def\otherfootline{\hfil}

\font\twelvebf=cmbx10 scaled\magstep 1
\font\twelverm=cmr10 scaled\magstep 1
\font\twelveit=cmti10 scaled\magstep 1

\font\tenbf=cmbx10

\parindent=1.5pc
\hsize=6.0truein
\vsize=8.5truein
\nopagenumbers

\centerline{\tenbf  REAL-TIME PROPAGATOR IN THE}
\baselineskip=22pt
\centerline{\bf FIRST QUANTISED FORMALISM${}^*$}
\baselineskip=16pt
\vglue 0.8cm
\bigskip
\bigskip

\centerline{\twelverm SAMIR D. MATHUR}
\baselineskip=13pt
\bigskip
\bigskip
\centerline{\twelveit  Center For Theoretical Physics,  M.I.T. , }
\baselineskip=12pt
\centerline{\twelveit  Cambridge, MA 02139, U.S.A.}
\vglue 3cm
\vglue 0.8cm
\centerline{\twelverm ABSTRACT}
\vglue 2cm
{\rightskip=3pc
 \leftskip=3pc
 \twelverm\baselineskip=12pt\noindent
We argue that a basic modification must be made to the first quantised
formalism of string theory if the physics of `particle creation' is to
be correctly described. The analogous quantisation of the relativistic
particle is performed, and it is shown that the proper time along the
world
line must go both forwards and backwards (in the usual quantisation
it only goes forwards). The matrix propagator of the real time formalism
is
obtained from the two directions of  proper time.

\vglue 0.6cm}

\vfil
\twelverm
\baselineskip=14pt
${}^*$ Talk given at the  Thermal Fields Workshop held
at Banff, Canada (August 1993)
\eject
\leftline{\twelvebf 1. The Issue}
\vglue 0.3cm
\vglue 1pt
One of the most well known problems concerns the ultimate fate of an
evaporating black hole. A common viewpoint is that this question can
only be
settled in a complete theory of quantised gravity plus matter. Strings
provide
such a theory; further, black hole solutions have been constructed for
strings${}^1$. Why then don't we have a complete understanding of
the fate of
the black hole?

One problem seems to be that it is not easy to get the string black hole
to
evaporate. Topological methods can  study the black hole to all string
loop
orders${}^2$ (evaporation should appear from one loop onwards). The
fact that
the black hole doesn't appear to evaporate has led to the speculation
that the
functional measure for matter fields is such for the string case that
there is
indeed no evaporation. Matrix model techniques should  also describe
the string
theory to all loop orders, but again there is no clear evaporating
solution.

In this talk we argue that there is a more basic problem encountered in
studying any process of particle (string) creation using the techniques
used in
studying strings. Strings are studied in the first quantised language.
For the
simpler but analogous case of a free scalar field theory, this
corresponds to
writing the propagation amplitude between two space-time points as
the sum over
all trajectories of a relativistic scalar particle travelling between those
two
points. Such a computation generates a propagator between the `in'
and the
`out' vacuua${}^3$:
$${D(x',x)=\int D[{\rm paths}({x\rightarrow x'})]e^{iS}
={}_{out}<0| T[\phi(x')\phi(x)]|0>_{in} /{}_{out}<0|0>_{in} }\eqno(1)$$
Using such a propagator in the string case (and enforcing the
vanishing
$\beta$-function conditions) leads to  Einstein equations incorporating
backreaction of created particles in the form${}^4$
$${G_{\mu\nu}~=~{
{}_{out}<0|T_{\mu\nu}|0>_{in}\over {}_{out}<0|0>_{in}}}\eqno(2)$$
But what we need on the RHS of Einstein's equations is a true
expectation value
of $T_{\mu\nu}$ in the physical state, not an `in-out' scattering
amplitude. In
fact when the `in' and `out' vacuua are not the same,  perturbation
theory
needs the $2x2$ matrix propagator of the real-time formalism.  Thus if
particle
creation and backreaction are to be correctly studied in the first
quantised
language used for strings then we must be able to find in a natural way
the
matrix propagator from the `sum over paths'  formalism. We will show
how this
emerges naturally with a careful quantisation of the relativistic scalar
particle, and will then comment on the string case.

\vglue 0.6cm
\leftline{\twelvebf 2. Quantising the Relativistic Particle}
\vglue 0.4cm
The geometric action for a scalar particle is
$${S~=~\int_{X_i}^{X_f} mds~=~\int_{X_i}^{X_f} m(X^\mu,_\tau
X_\mu,_\tau )^{1/2}d\tau~\equiv~\int L d\tau}\eqno(3)$$
 where $\tau$ is an arbitrary
parametrisation of the world line. The canonical momenta
$${P_\mu~=~{\partial L\over \partial X^\mu,_\tau}~=~{mX_\mu,_
\tau\over (X^\mu,_\tau X_\mu ,_\tau)^{1/2}}}\eqno(4)$$
satisfy the constraints
$${P^\mu P_\mu-m^2~=~0}\eqno(5)$$
We choose the range of the parameter $\tau$ as $[0,1]$. Following the
approach  given in Kaku${}^5$, we impose the constraint at each
$\tau$ through
a
$\delta$-function:
$${\delta(p^2(\tau)-m^2)~=~{1\over 4\pi}\int_{-\infty}^\infty
d \lambda(\tau)e^{-i\lambda/2(p^2(\tau)-m^2)}} \eqno(6)$$
The path integral amplitude
to propagate from $X_i$  to $X_f$ becomes
$${G(X_2,X_1)~\equiv~
N\int{D[X]D[P]D[\lambda]\over{\rm Vol}[{\rm Diff}]}
e^{i\int_0^1 d\tau[P_\mu(\tau)
X^\mu,_\tau(\tau)-\lambda/2(\tau)(p^2(\tau)-m^2)]}}\eqno(7)$$
 where $N$ is a
normalisation constant, $P_\mu X^\mu ,_{\tau}=m(X^\mu,_\tau X_\mu
,_\tau)^{1/2}$ is the original Lagrangian in Eq.~(3)
and we have divided by
the volume of the symmetry group, which which is related to
$\tau$-diffeomorphisms in the manner discussed below. (The
$\delta$-function constraint on the momenta and dividing by ${\rm
Vol}[{\rm Diff}]$ remove the two phase space co-ordinates redundant
in the
description of the particle path.)

The action is invariant under
$${\eqalign{{\cal S}: \quad\quad&\delta
X^\mu (\tau)~=~\epsilon(\tau)\lambda(\tau) P^\mu(\tau) \cr &\delta
P_\mu(\tau)~=~0 \cr &\delta
\lambda(\tau)~=~(\epsilon(\tau)\lambda(\tau)),_\tau \cr}} \eqno(8)$$
$\lambda$ transforms as an einbein under the
diffeomorphism $\tau\rightarrow \tau '(\tau)$. Note that for regular
$\epsilon(\tau)$, $\lambda$ either changes sign for no $\tau$ or for all
$\tau$. We take ${\rm Diff}$ as the group of regular diffeomorphisms
connected to the identity; then $\lambda$ does not change sign under
the allowed diffeomorphisms.   These
diffeomorphisms cannot gauge-fix $\lambda(\tau)$ to any preassigned
function $\lambda_1(\tau)$. $\lambda(\tau)$ and   $\lambda_1(\tau)$
must
have the same value of
$${\int_0^1\lambda(\tau)d\tau~\equiv~\Lambda}\eqno(9)$$
$\Lambda$ may  be interpreted as the length of
the world line.
 The restriction Eq.~(9) is usually assumed to mean that the
length of the world line is the only remaining parameter after
gauge-fixing. But what we find instead is that there is a
further complication: there is a discrete infinity of
classes, each with one or more continuous parameters.  One
class comes from configurations $\lambda(\tau)$ which are
everywhere
positive. This class can be gauge-fixed  with the
allowed diffeomorphisms to have
$${\dot\lambda(\tau)~=~0, \quad\quad \int_0^1 d\tau \lambda
(\tau)~=~\Lambda }\eqno(10)$$
 with $0<\Lambda<\infty$. Similarily, the set of
everywhere negative $\lambda(\tau)$ can be gauge-fixed as in
Eq.~(10) but
with $-\infty<\Lambda<0$.
Thus these classes give for the Fourier transform of  Eq.~(7)
$${\eqalign{G_{\Lambda>0}(p)~=&~
{1\over 4\pi}\int_{0}^\infty
d \lambda(\tau)e^{-i\lambda/2(p^2(\tau)-m^2)}~=~
{i\over p^2-m^2+i\epsilon},\cr
G_{\Lambda<0}(p)~=&~
{1\over 4\pi}\int_{-\infty}^0
d \lambda(\tau)e^{-i\lambda/2(p^2(\tau)-m^2)}~=~
{-i\over p^2-m^2-i\epsilon}\cr}}\eqno(11)$$
respectively. (We have added the $\epsilon$ term for
convergence to each sector; this term was not in the action.)
 Keeping the first class alone gives the Feynman
propagator for particles, while the second gives its complex
conjugate.

But in the path integral over $\lambda$ in  Eq.~(7)
we also
have the class of  $\lambda(\tau)$ which are positive for
$0<\tau<\tau_1$,
negative for
$\tau_1<\tau<1$. The group of orientation preserving diffeomorphisms
can
gauge fix this to
$${\dot\lambda(\tau)=0 ~~{\rm for} ~~\tau\ne\tau_1,
{}~~~~\int_0^{\tau_1} d\tau \lambda(\tau)=\Lambda_1,
{}~~\int_{\tau_1}^1
d\tau
\lambda(\tau)=\Lambda_2}\eqno(12)$$
with $0<\Lambda_1<\infty$, $-\infty<\Lambda_2<0$. We would like to
identify this sector as the contribution to the amplitude to start with a
state of type $1$ and end with a state of type $2$ (the off-diagonal
element $D_{12}$ of the matrix propagator).  A
general sector has a given number of alternations
in the sign of $\lambda$, and in each interval of
constant sign we gauge fix $\lambda$
to a constant $\Lambda_i$.  We can perform the integration
over the variables $\Lambda_i$ appearing in each sector,  but
we should also specify a `transition amplitude' for each point $\tau_i$
where $\lambda$ changes sign.  We allow this
 amplitude to depend on the
states on both sides  of $\tau_i$, and also on whether
$\lambda$ changes from positive to negative or vice versa.
 This amlplitude is not supplied by the original action;
it is supplementary information needed for determing the propagator.

We will argue with examples below that we should identify
$\lambda>0$ with
particles of the type 1
field  and $\lambda<0$  with  particles of the
type two field in the language of  the real time formalism for
perturbation
theory.  To obtain the matrix
propagator element $D_{ji}$ ($i,j=1,2$)  we need to  add together all
sectors
for $\lambda(\tau)$ beginning as type
$i$ and ending as type $j$. When the einbein $\lambda$ changes sign
we may say
that the proper time along the world line has reversed direction.

\vglue 0.6cm
\leftline{\twelvebf 3. Examples}
\vglue 0.4cm
First we note another way of arriving at the proper time representation
of the propagator.
 The Feynman propagator  for a scalar field in Minkowski space can be
written as
$${D_F(p)~=~{i\over p^2-m^2+i\epsilon}~=~\int_{\tilde
\lambda=0}^\infty d \tilde \lambda e^{i\tilde\lambda
(p^2-m^2+i\epsilon)}}\eqno(13)$$
Eq.~(13) can be used to express $G_F(p)$ in a first quantised
language,
with $p^2=-\boxx$.  The Hamiltonian on the world line
is $\boxx+m^2$, evolution takes place in a fictitious time for a duration
$\tilde \lambda$, and this length $\tilde \lambda$ of the world line is
summed over all values from $0$ to $\infty$.

Now consider the scalar field in Minkowski space at temperature
$\beta^{-1}$.
For the closed time path beginning and ending at $t=-\infty$ the matrix
propagator is
$${D(p)~=~\pmatrix{{i\over
p^2-m^2+i\epsilon}+2\pi n(p)\delta(p^2-m^2) &
2\pi[(n(p)+\theta(-p_0)]\delta (p^2-m^2) \cr 2\pi[n(p)+\theta(p_0)]
\delta(p^2-m^2) & {-i\over p^2-m^2-i\epsilon}+2\pi
n(p)\delta(p^2-m^2)\cr}}
\eqno(14)$$
($n(p)$ is the number density.) We can write this as
$${D(p)~=~ \int_0^\infty d\tilde \lambda
e^{-i\tilde\lambda H-\epsilon\tilde\lambda M},~~H=\pmatrix
 {-(p^2-m^2)&0\cr 0&(p^2-m^2)\cr }}$$
$$ M=
\pmatrix{1+2n(p)&-2\sqrt{n(p)(n(p)+1)}\cr -2\sqrt{n(p)(n(p)+1)}&
1+2n(p)\cr}\eqno(15)$$
Comparing with Eq.~(13) we observe the following. To obtain the
matrix propagator in a many-body situation  we need to consider both
evolutions
$e^{-i\tilde\lambda H}$ and $e^{i\tilde \lambda H}$ on the world line. A
regulating  factor is needed to define the first quantised path integral,
near the mass shell. But the regulator matrix $M$ need not be
diagonal in the
two kinds of world line evolution.

We can now relate the world line evolution in Eq.~(15) to the general
discussion of  proper time quantisation in the previous section.  The
two
propagations $\pm(i\tilde\lambda (p^2-m^2)$ correspond to the two
signs of
$\lambda$ that arose in the gauge fixing of the relativistic particle.
We find that  to obtain Eq.~(14) from  the proper time quantisation of
the
relativistic particle we  must take the amplitude for $\lambda$ to
 go from positive to negative as $\tilde\epsilon {\rm sech}(\beta|p_0|/2)
e^{\beta p_0/2}$, and for negative to positive as
 $\tilde\epsilon {\rm sech}(\beta|p_0|/2)e^{-\beta p_0/2}$. Indeed,
extending
the computaion of Eq.~(11) to all `sectors' arising in the gauge fixing,
and
summing over sectors with the above amplitudes for orientation
reversal, we
reproduce the matrix propagator Eq.~(14)${}^4$.

As another example, consider the free scalar field propagating in
$1+1$ spacetime with metric
$${ds^2~=~C(\eta)[d\eta^2-dx^2], \quad\quad -\infty<\eta<\infty,
\quad 0\le x<2\pi}\eqno(16)$$
$${C(\eta)~=~A+B\epsilon(\eta), \quad\quad A>B\ge 0}\eqno(17)$$
This describes a Universe with `sudden' expansion, so we expect
particle creation. Let the physical state be the `in' vacuum $|0>_{in}$.
Then we need the matrix propagator
$${D(z_2,z_1)~=~\pmatrix{{}_{in}
<0|T[\phi(z_2)\phi(z_1)]|0>_{in} & {}_{in}<0|\phi(z_1)\phi(z_2)|0>_{in}\cr
{}_{in} <0|\phi(z_2)\phi(z_1)|0>_{in}& {}_{in}<0|\tilde
T[\phi(z_2)\phi(z_1)]|0>_{in}\cr}}
\eqno(18)$$
Our goal is to see if there exists a choice of regulator matrix
$M$ such that evaluting
$${D~=~ \int_0^\infty  \lambda
e^{-i\lambda H-\epsilon\lambda M}, \quad  H=\pmatrix
 {(\boxx+m^2)&0\cr 0&-(\boxx+m^2)\cr }}\eqno(19)$$
gives the matrix propagator Eq~(18).
Here $\boxx$ acts on functions on spacetime $f(\eta,x)$.

Fourier modes  $\sin(nx)$, $\cos(nx)$ of the scalar field  decouple from
each other.  Consider for one such mode the basis of  solutions
of the field equation
$${\eqalign{f^1_n~=~&{\cos(nx)\over
\sqrt{2\pi}}e^{-i\omega_n^- \eta}, \quad \quad \eta<0\cr
=~&{\cos(nx)\over \sqrt{2\pi}}[{1\over 2}(1+{\omega_n^-\over
\omega_n^+})
e^{-i\omega_n^+\eta} ~+~ {1\over 2}(1-{\omega_n^-\over
\omega_n^+})e^{i\omega_n^+ \eta}],\quad\eta>0 \cr
f_n^2(\eta,x)~=~&f_n^{1*}(\eta,x) \cr }}\eqno(20)$$
($\omega_n^\pm$ are the frequencies of the mode for $\eta>0$,
$\eta<0$
respectively.)  In the flat space example above we had space-time
translation
invariance, so the `regulator matrix' $M$ was diagonal in the
4-momentum $p$.
In the present example
 there is no translational invariance in  $\eta $, so
the modes $f_n^1$, $f_n^2$ can `connect' across the point on the
world line
where the einbein $\lambda$ changes sign, in the proper time path
integral. Thus for each $n$ we need to consider a 4x4 matrix, which
acts on a
column vector $(\{f_{n}^1, f_{n}^2\}^+,\{f_{n}^1,f_{n}^2\}^-) $.
Here the first pair of functions propagate on the world line as $e^{i
(\boxx+m^2)}$
while the second pair propagates as $e^{-i(\boxx+m^2)}$.  The result
we get is
that the matrix  propagator  Eq.~(18) is obtained for
$${M~=~\pmatrix{1&0&0&0\cr 0&1&{4B_n\over 1+4B_n^2}&
{-2\over 1+4B_n^2}\cr {-2\over 1+4B_n^2}&{4B_n\over 1+4B_n^2}&
1&0&\cr0&0&0&1\cr}}\eqno(21)$$
 where
$${B_n~=~-{1\over 2}{\omega_n^+-\omega_n^-\over
\omega_n^++\omega_n^-} }\eqno(22)$$
($B_n$ are proportional to the Bogoliubov co-efficients.)
If we had not allowed for the reversal of the proper time direction (i.e.
$M$
was the identity matrix) then we would get the `in-out' propagator
Eq.~(1)
instead of Eq.~(18).

\vglue 0.6cm
\leftline{\twelvebf 4. Strings}
\vglue 0.4cm
We show that the real time matrix propagator is `natural' for
the first quantised string in the same sense that interactions
are natural for a string. For interactions we note that one closed string
can split into two closed strings (the pants diagram) with the world
sheet
smooth at all points. In a similar sense we can get a smooth analogue
of the
einbein sign change discussed above for the particle.  We have
a Minkowski signature target space, which requires a Minkowski
signature
world sheet. The world sheet metric and action are
$${ds^2=g_{ab}dx^adx^b,~~~~S=\int d^2x \sqrt{-g}g^{ab}
{X^\mu,}_aX_{\mu,b}}\eqno(23)$$
where $x^0=\tau$, $x^1=\sigma$ are the world sheet coordinates and
$X^\mu$ are
the target space (i.e. spacetime) coordinates. Consider the metric
$${g_{ab}=\pmatrix{-\tanh \tau & {\rm sech} ~\tau \cr {\rm sech} ~\tau &
\tanh\tau\cr},~~~~\sqrt{-g}=1}\eqno(24)$$
$${\eqalign{\tau\rightarrow -\infty,~~&g_{ab}\rightarrow
{\rm{diag}}\{1,-1\},~~S\rightarrow \int \partial_\tau X^\mu\partial_\tau
x_\mu
-\partial_\sigma X^\mu\partial_\sigma X_\mu \cr
\tau\rightarrow \infty,~~&g_{ab}\rightarrow
{\rm{diag}}\{-1,1\},~~S\rightarrow \int (-)\partial_\tau X^\mu\partial_\tau
x_\mu
+\partial_\sigma X^\mu\partial_\sigma X_\mu \cr}}\eqno(25)$$

In the particle limit for the string, $\partial_\tau X^\mu$ gives the
${X^\mu,}_\tau$ for the particle case and $\partial_\sigma X^\mu$
gives the mass to the particle. Thus in the limits
$\tau\rightarrow\pm\infty$,
$S$ corresponds to the Hamiltonians $\mp (p^2-m^2)$ on the
particle world line, and we reproduce the `orientation change' along the
world
line with the string world sheet metric everwhere regular
($\sqrt{-g}=1$).
Since we have to sum over all geometries on the world sheet we are
forced to
consider metrics such as Eq.~(24).  It is appropriate that in a theory of
gravity (i.e. string theory) we are naturally led to the real time
propagator
because
generic solutions of such a theory give curved space with inequivalent
`in' and
`out' vacuua and consequent particle creation.

We note that the field equations for  the background spacetime in
string theory
are derived not from a Lagrangian but from requiring consistent
propagation of
the string in the  background.  It is the causal `in-in' propagator that
must
be `consistent', not the `in-out' vacuum propagator, if we are to get the
correct backreaction equations and not  an equation like Eq.~(2).  We
have
shown how the `first quantised' language can obtain the causal
propagator, and
moreover that the required `orientation reversal' is natural to strings.

Our central point, viz. that proper time need not go only forwards,
may  have relevance beyond the first quantization approach to strings.
The quantization of gravity is very much like the quantization of the
relativistic particle. In  the canonical  approach, the wavefunction
depends
on 3-geometries,  and a semiclassical 4-geometry emerges only if
one makes a  WKB  ansatz for the dominant mode of the 3-metric. The
`time'
direction so obtained looks continuous presumably only at scales
much larger than one period of the WKB waveform. Thus  the
semiclassical nature
of the 4-geometry would break down when the geometry is examined
at
sufficiently small scales. This can happen for instance when we try to
relate
the Kruskal co-ordinates to the Schwarzschild co-ordinates at the
horizon of a
black hole. One might think that the problem would be avoided in a
path-integral formalism using a continuous proper time to describe the
4-geometries, but here one should consider the possibility that the
proper time
may not run only forwards. Issues
like the third quantization of geometries  then appear in the same way
that we
obtained the effect of many particles (second quantization ) from the
study of just one relativistic trajectory.

\vglue 0.6cm
\leftline{\twelvebf 5. Acknowledgements}
\vglue 0.4cm
This work is supported in part by DOE grant DE-AC02-76ER.

\vglue 0.6cm
\leftline{\twelvebf 6. References}
\vglue 0.4cm

\medskip
\itemitem{1.} E. Witten,  {\twelveit Phys. Rev.} {\twelvebf D44} (1991)
314.
\itemitem{2.} S. Mukhi and C. Vafa {\twelvebf HUTP-93-A002} (Jan
1993).
\itemitem{3.} H. Rumpf and H.K.
Urbantke, {\twelveit Ann. Phys.} {\twelvebf 114} (1978) 332,  H. Rumpf,
{\twelveit Phys. Rev.} {\twelvebf D 24}
(1981) 275, {\twelvebf D 28} (1983) 2946.
\itemitem{4.} S.D. Mathur {\twelvebf hep-th 9306090}.
\itemitem{5.} M. Kaku, {\twelveit Introduction to
Superstrings} (Springer-Verlag, 1988).

\bye